\documentclass{jfm}
\usepackage{graphicx}
\usepackage{epstopdf}
\usepackage{natbib}
\usepackage{hyperref}
\usepackage{natbib}
\usepackage{amsmath,mathtools}
\usepackage{graphicx}
\usepackage{adjustbox}
\hypersetup{
    colorlinks = true,
    urlcolor   = blue,
    citecolor  = black, 
}
\usepackage{color}
\usepackage{bm}
\usepackage{subcaption} 
\usepackage{enumitem} 
\newcommand{\RomanNumeralCaps}[1]
\linenumbers

\def\beq{\begin{equation}}
\def\eeq{\end{equation}}
\newcommand{\per}{\, .}
\newcommand{\com}{\, ,}

\newcommand{\Ac}{\mathcal{A}}

\newcommand{\Pc}{\mathcal{P}}

\newcommand{\tha}{\theta}

\newcommand{\defn}{\stackrel{\text{def}}{=}}

\newcommand{\rv}{r_{\mathit{v}}}

\def\ee{\mathrm{e}}
\def\ii{\mathrm{i}}
\def\bx{{\bm{x}}}

\def\bk{{\bm{k}}}
\def\bU{\bm{U}}

\def\bq{\bm{q}}
\def\bp{\bm{p}}

\def\bP{\bm{P}}
\def\eps{\varepsilon}
\def\dd{\mathrm{d}}

\def\bHs{\bar {H}_s}
\def\bA{\bar \Ac}
\def\Hs{H_s}
\def\hs{h_s}
\def\ha{\hat a}
\def\hbU{\hat{\bU}}
\def\hhs{\hat{h}_s}
\def\bL{\bm{L}}
\def\hbl{\hat{\bm{L}}}
\def\hlpar{\hat{{L}}_{\parallel}}
\def\hlper{\hat{{L}}_{\perp}}
\def\eq{\bm{e}_{\bm{q}}}
\def\eqp{\bm{e}_{\bm{q}}^\perp}

\def\qang{\varphi}
\def\xang{\nu}

\renewcommand\Re{\mathop{\mathrm{Re}}}
\renewcommand\Im{\mathop{\mathrm{Im}}}

\newcommand{\dt}[2]{\frac{\dd #1}{\dd #2}}

\definecolor{HW}{RGB}{137,0,225}

\def\Xint#1{\mathchoice
   {\XXint\displaystyle\textstyle{#1}}%
   {\XXint\textstyle\scriptstyle{#1}}%
   {\XXint\scriptstyle\scriptscriptstyle{#1}}%
   {\XXint\scriptscriptstyle\scriptscriptstyle{#1}}%
   \!\int}
\def\XXint#1#2#3{{\setbox0=\hbox{$#1{#2#3}{\int}$}
     \vcenter{\hbox{$#2#3$}}\kern-.5\wd0}}

\def\dashint{\Xint-}

\newcommand\ch[1]{#1}

\def\bc{\bm{\ch{c_g}}}

\title{Scattering of  surface waves by ocean currents: the U2H map}

\author{Han Wang\aff{1} \corresp{\email{hannnwangus@gmail.com}}, 
Ana~B. Villas B\^oas\aff{2}, 
Jacques Vanneste\aff{1}
  \and
  William R. Young\aff{3}}

\affiliation{\aff{1}School of Mathematics and Maxwell Institute for Mathematical Sciences, University of Edinburgh, EH9 3FD, UK
 \aff{2}Department of Geophysics, Colorado School of Mines, Golden CO 80401, USA
 \aff{3} Scripps Institution of Oceanography, University of California at San Diego, La Jolla CA 92093-0213, USA
 }

\begin{document}
\maketitle

\begin{abstract}
Ocean turbulence at meso- and submesocales affects the propagation of surface waves through refraction and scattering, inducing spatial modulations in significant wave height (SWH). We develop a theoretical framework that relates these modulations to the current that induces them. 

We exploit the asymptotic smallness of the ratio of typical current speed to wave group speed to derive a linear map -- the U2H map -- between surface current velocity and SWH anomaly. The U2H map is a convolution, non-local in space, expressible as a product in Fourier space by a factor   independent of the magnitude of the wavenumber  vector.  Analytic expressions of the U2H map show how the SWH responds differently to the vortical and divergent parts of the current, and how the anisotropy of the wave spectrum is key to large current-induced SWH anomalies. 

We implement the U2H map numerically and test its predictions against  WAVEWATCH III numerical simulations for both idealised and realistic current configurations. 

\medskip

\end{abstract}

%
%

\section{Introduction}
\label{sec:intro}

Surface gravity waves (SGWs)  propagate through  currents resulting from ocean  \ch{meso- and} submesoscale turbulence   and from  the surface expression of tides and internal gravity waves.  Much  work, both historical and recent,  has focussed on the effect of  internal waves and tides on SGWs  \citep[e.g.][]{Barber1949,PS1965,Phillips1977, OB1980,Tolman1990, Hao2020,Ho2023}. Recognition of the role of \ch{meso- and} submesoscale turbulence in shaping the open-ocean surface wave field is comparatively recent  and relies on  ocean observations and modeling \ch{\citep{holthuijsen1991effects, Romero2017, Ardhuin2017,Romero2020,VillasBoas2020}}.  
In this work we  develop a theoretical framework that can be used to understand the effect of \ch{meso- and} submesoscale turbulence  on  SGWs. 

\begin{figure}
  \centering
  \includegraphics[width=1.0\textwidth]{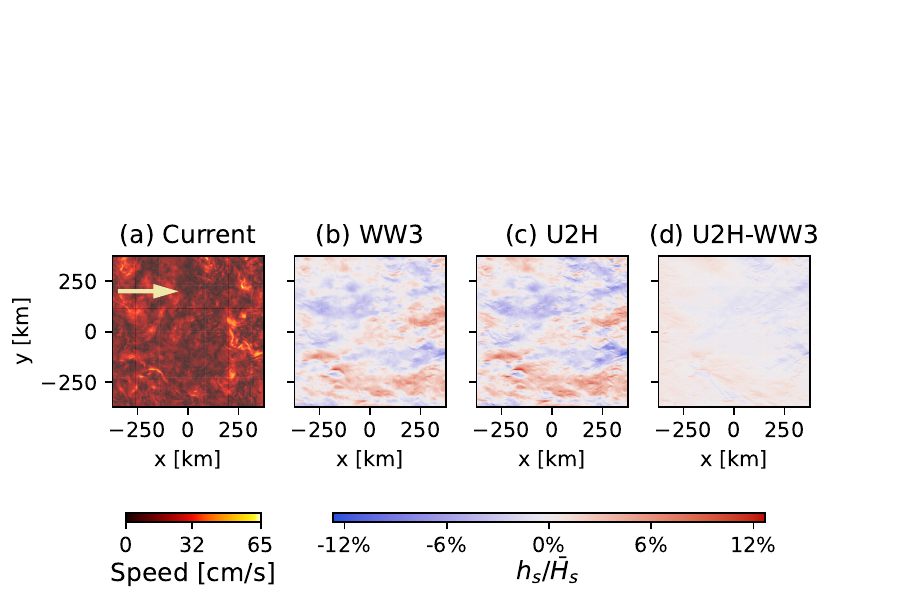}
    \caption{(a): surface current speed in an MITgcm simulation of the California Current system  \citep{VillasBoas2020}\ch{, with the arrow indicating the primary direction of wave propagation.} \ch{SWH anomaly}
computed using (b) WW3  and (c) the U2H map. \ch{(d) difference between (c) and (b).}  \ch{The wave spectrum,  described in Appendix \ref{app:LHCS},  is  narrow banded in frequency (with periods around 10.3 s and wavelength 165.5 m) and  angle (with peak angle $\theta_p=0$ and width parameter $s=10$).}
Panels (a) and (c) can be produced from the notebook  accessible at \url{https://shorturl.at/bef14}.}
  \label{fig:hs_mitgcm}
\end{figure}

Refraction and scattering of open-ocean deepwater SGWs by  eddies, fronts, filaments  and vortices  results in fluctuations in significant wave height  (SWH) with length  scales  reflecting  those of the underlying  turbulent field -- see figure \ref{fig:hs_mitgcm}. Fluctuations in SWH modulate SGW breaking and thus affect all aspects of  air-sea exchange \citep{Cavaleri2012,Boas2019}. There are also implications for the  formation of  extreme waves and remote sensing.

Scale separation between SGW  wavelengths and the larger spatial scale of the currents makes it possible to adopt a phase-averaged description, focusing on the density of wave action $\mathcal{A}(\bx,\bk,t)$ 
in the position--wavevector $(\bx,\bk)$-space. Action density satisfies a transport equation which, because of its high dimensionality, poses analytic and numerical challenges, even when linearised by neglecting wave--wave interactions. As a result, numerous open questions remain about the relation between the currents and the fluctuations in SWH they induce. 
Some of these questions can be addressed by numerical solution of the transport equation \citep{VillasBoas2020}. These computations are costly and the  results  can be difficult to interpret.

 In \S\ref{sec:formulation} we develop an alternative approach that directly links SGW amplitude to current,  reducing computational costs and providing new  insights. This approach relies on the smallness of the ratio {$\eps$ between the typical current speed $U$ and the SGW group speed $\ch{c_g}$: }
 \ch{
 \beq \label{eps}
 \eps = U/\ch{c_g} \ll 1.
 \eeq
 }
SGWs with wavelengths greater than  10 m have  group speed in excess of 2 m s$^{-1}$. For these relatively long waves  $\eps \ll 1$ holds in all but the most extreme ocean conditions.
In a steady-state scenario and in the absence of current, that is, for $\eps = 0$, the action density $\Ac(\bx,\bk)$  can be taken as spatially uniform, $\bar \Ac (\bk)$ say.  For $0 < \eps \ll 1$ \ch{and neglecting wind forcing, dissipation and wave-wave interactions}, the current induces a small, $O(\eps)$ $\bx$-dependent modulation so that the current-perturbed action spectrum is  $\Ac(\bx,\bk)= \bar\Ac(\bk)+a(\bx,\bk)$. 
\ch{At  leading order, the} anomaly $a(\bx,\bk)$  is linearly related to \ch{both} the surface current \ch{$\bU(\bx)$ and the background wave action $\bar\Ac$}.
Similarly the anomaly of any measure of wave amplitude deduced from $\Ac(\bx,\bk)$ is linearly related to \ch{$\bU(\bx)$  and $\bar \Ac(\bk)$}. 

We focus on SWH and obtain an explicit form for the linear map that relates the  SWH anomaly $\hs(\bx)$ to the current velocity $\bU(\bx)$. This map, which we term U2H map, turns out to be a convolution, best expressed as
\beq
\frac{\hhs}{\ch{\bHs}}= \hbl \bcdot \hbU
\label{U2H0}
\eeq
in terms of the Fourier transforms \ch{$\hhs$ of $\hs$ and $\hbU$ of $\bU$, \ch{and of the mean SWH $\bHs$.}
}  
In section \ref{sec:U2H} we obtain several alternative and approximate  expressions  for the transfer function $\hbl$ that embodies the U2H map. Depending on details of the   currents $\bU$ and  the base-state action spectrum $\bar \Ac(\bk)$ one of these  different expression of $\hbl$ may be most effective.

We show a complicated example in figure \ref{fig:hs_mitgcm}. Panel (a) shows the surface current speed in a simulation of the California Current system. Panel (b) shows the SWH anomaly $\hs$ computed using  WAVEWATCH III (Wave Height, Water Depth, and Current Hindcasting third generation wave model, hereafter WW3) which solves the four-dimensional transport equation satisfied by the action $\Ac(\bx,\bk,t)$ \citep{Tolman2009user}. Panel (c) shows the result of applying the U2H map to the current in panel (a). \ch{The computational details for panels (b) and (c) are described in \S \ref{sec:numimp}.} The match between the results of the (computationally expensive) WW3 computation in (b) and of the (much cheaper) application of U2H in (c) is excellent. Throughout the paper we assess the accuracy of the U2H map \eqref{U2H0} by comparing its predictions to numerical simulations using WW3.

In  \S\ref{sec:particularflows} we examine the SWH anomaly $\hs$ induced by \ch{realistic flows and by} simple flows  such as vortices. In  \S\ref{sec:particularspectra}  we consider $\hs$ produced by  special wave spectra. We show, in particular, that $\hs$ vanishes (to leading order in $\eps$) for isotropic wave spectra. The complementary  limit is highly directional wave spectra, characteristic of ocean swell. Swell  produces strongly anisotropic SGW anomalies aligned with the dominant direction of wave propagation, i.e.\ streaks in SWH. The limit of highly directional wave spectra is delicate in that it is non-uniform in the small parameter, $\delta$ say, characterising the \ch{directional} width. The results of the present paper \ch{require} that $ \delta \gg \eps$. For $\delta = O(\eps)$, the SWH response is nonlinear in the current and the assumptions leading to the U2H map break down. An asymptotic form for the SWH in this case is derived in \citet{WVBYV} under the additional assumption of a localised current.

\section{Formulation}\label{sec:formulation}

We start with the conservation equation
\beq
\partial_t \Ac + \bnabla_\bk  \omega \bcdot \bnabla_\bx \Ac - \bnabla_\bx \omega \bcdot \bnabla_\bk \Ac = 0
\label{actioncons}
\eeq
for the wave action density \ch{per unit mass} $\Ac(\bx,\bk,t)$  in  position--wavevector  space  \citep[e.g.][]{KomenBook,Janssen}. Here $\omega$ is the absolute frequency of deep-water SGWs, related to the intrinsic frequency 
\beq
\ch{\sigma(k) = \sqrt{gk}}
\eeq
with $k = | \bk|$, by
\beq
\omega(\bx,\bk) = \sigma(k) + \bm{k} \bcdot \bU(\bx),
\label{frequency}
\eeq
where $\bm{U}(\bx)$  is the surface velocity of the ocean current, taken to be horizontal and time independent.
The wave energy \ch{per unit mass} spectrum is related to the action density according to
\beq
\mathcal{E}(\bx,\bk,t)=\sigma(k) \Ac(\bx,\bk,t).
\label{energy}
\eeq
With equipartition between kinetic and potential energy of deep-water SGW the root mean square sea-surface displacement is related to the wave energy spectrum by 
\beq
g  \eta^2_{\mathit{rms}} = \int \mathcal{E}(\bx,\bk,t) \, \dd \bk. \label{re7}
\eeq

\ch{The polar representation of the wavenumber vector $\bk$ is 
\beq
\bk = k \begin{pmatrix}\cos \theta \\ \sin \theta\end{pmatrix},
\label{k1}
\eeq
and integrations in the form of $\int (\cdot)\dd \bk$ (as in \eqref{re7}) can be expressed as $\iint   (\cdot)  k\dd k \dd \theta$.
Compared to conventions adopted in wave modelling communities, our definitions of the wave action density per unit mass $\Ac(\bx,\bk,t)$ and wave energy  per unit mass spectrum  $\mathcal{E}(\bx,\bk,t)$  are related to the definitions of wave action density $N(k,\theta;\bx,t)$ and surface elevation spectrum $F(k,\theta;\bx,t)$  as appear in WW3 \citep{wavewatch2016user} via $\Ac(\bx,\bk,t)=gN(k,\theta;\bx,t)$ and $\mathcal{E}(\bx,\bk,t) = g F(k,\theta;\bx,t)$.
} 

Our focus is on the spatial distribution of wave energy, obtained by integrating \eqref{energy} in $\bk$ and conventionally reported in terms of the SWH defined as  $H_s(\bx,t) = 4 \eta_{rms}(\bx,t)$.  SWH  can be obtained from the action spectrum with 
\beq
H_s(\bx,t) = 4 \left( g^{-1} \int \sigma(k) \Ac(\bx,\bk,t) \, \dd \bk \right)^{1/2}\per 
\label{swh}
\eeq

The action equation  \eqref{actioncons} relies on an assumption of spatial scale separation between surface gravity waves and currents. \ch{It also neglects forcing, dissipation and wave--wave interactions. We make three further assumptions: 
\begin{enumerate}
\item  \  in the absence of currents, the wave action takes a background value $\bA(\bk)$ that is independent of space and time; 
\item \ the current is steady and we restrict our attention to the steady-state wave-action response; 
\item \   the typical current velocity is small compared with the typical group speed of SGWs,  so that the Doppler shift  $\bm{k} \bcdot \bU(\bx)$ is a small correction to the intrinsic frequency $\sigma(k)$. 
\end{enumerate}
}

\ch{
As preparation for a perturbation expansion  we make assumption \textit{(iii)} explicit by writing
\beq
\omega(\bx,\bk) = \sigma(k) + \eps \bm{k} \bcdot \bU(\bx).
\label{freq7}
\eeq
Here we avoid a formal scaling analysis and retain dimensional variables. Thus $\eps$ should from now on be regarded as a bookkeeping parameter that identifies terms that are $O(U/c_g)$ and  is ultimately set to $1$. We
}
expand the action as 
\beq
\Ac(\bx,\bk,t) = \bA(\bk) + \eps a(\bx,\bk,t) + O(\eps^2),
\label{exp}
\eeq
assuming that the leading-order action $\bA(\bk)$ is independent of space and time.
The presence of the currents leads to anomalies at order $\eps$ captured by $a(\bx,\bk,t)$.  We relate $a(\bx,\bk,t)$ to $\bU(\bx)$ by introducing \eqref{exp} into \eqref{actioncons} to obtain
\beq
\partial_t a + \bc \bcdot \bnabla_{\bx}   a =  (\bnabla_{\bk} \bA \bcdot \bnabla_\bx) \bU \bcdot \bk,
\label{a_t}
\eeq
where $\bc(k)= \sqrt{g/4k^3} \bk$ is the group velocity. We focus on the steady-state response $a(\bx,\bk)$, independent of $t$. This satisfies 
\eqref{a_t} where the time derivative term is omitted. Causality is enforced by adding a linear dissipation term to find
\beq
(\bc \bcdot \bnabla_{\bx} + \mu) a =  (\bnabla_{\bk} \bA \bcdot \bnabla_\bx) \bU \bcdot \bk.
\label{a_t2}
\eeq
The non-dissipative, causal solution is then obtained in the limit $\mu \to 0^+$.

We solve \eqref{a_t} in terms of the Fourier transforms 
\beq
\ha(\bq,\bk)  \defn \int  a(\bx,\bk) \, \ee^{-\ii \bq \cdot \bx} \, \dd \bx  \quad \textrm{and} \quad 
\hbU(\bq)  \defn \int \bU(\bx) \, \ee^{-\ii \bq \cdot \bx} \,  \dd \bx.
\label{fourier}
\eeq
We emphasise the distinction between the newly introduced wavevector $\bq$, which captures spatial variations at the current scale, and the original wavevector $\bk$, which represents spatial variations of the wave phase. Introducing \eqref{fourier} into \eqref{a_t2} leads to
\beq
\ha(\bq,\bk) = \lim_{\mu \to 0^+} \frac{(\bk \bcdot \hbU) (\bq \bcdot \bnabla_{\bk} \bA)}{\bc \bcdot \bq - \ii \mu}.
\label{fourier7}
\eeq
The limit $\mu \to 0^+$  above  is taken in all expressions involving $\mu$ and  we proceed  with lighter notation in which the limit is understood.

Our focus is on the SWH, which we expand as
\beq
H_s(\bx) = \bHs + \eps \hs(\bx) + O(\eps^2).
\label{Hsexp}
\eeq
The leading order term $\bHs$ is a constant. The anomaly $\hs(\bx)$ is deduced from $a(\bx,\bk)$ by
Taylor expanding \eqref{swh}:
\beq
\hs(\bx) = \frac{8}{g \bHs} \int \sigma(k) a(\bx,\bk) \, \dd \bk \per
\label{hsa}
\eeq
An analogous formula relates the Fourier transform $\hhs(\bq)$ of $\hs(\bx)$ to $\ha(\bq,\bk)$.
Substituting \eqref{fourier7} into \eqref{hsa} gives
\beq
\ch{\frac{\hhs(\bq)}{\bHs}} = \hbl(\bq) \bcdot \hbU(\bq)\com
\label{linmap}
\eeq
where
\beq
\hbl(\bq) = \frac{8}{g \ch{\bHs^2}}   \int\frac{ \bq \bcdot \bnabla_{\bk} \bA}{\bc \bcdot \bq - \ii \mu} \sigma \bk \, \dd \bk.
\label{bL1}
\eeq

Eq.\ \eqref{linmap} shows that $\hs$ is obtained from $\bU$ via a linear map -- the U2H map. This map is naturally expressed in terms of Fourier transforms, with the complex vector $\hbl(\bq)$ acting as a transfer function. It is clear from \eqref{bL1} that $\hbl$ depends on the wavevector $\bq$ only through its orientation: introducing the polar representation 
\beq
\bq = q \, \eq \defn q \begin{pmatrix}\cos \qang\\ \sin\qang\end{pmatrix}\com
\label{q1}
\eeq
with $\eq$ the unit vector in the direction of $\bq$ and $-\pi < \qang \le \pi$, we can rewrite \eqref{bL1} as
\beq
\hbl(\bq) = \hbl(\qang) = \frac{8}{g  \ch{\bHs^2}}   \int\frac{ \eq \bcdot \bnabla_{\bk} \bA}{\bc \bcdot \eq - \ii \mu} \sigma \bk \, \dd \bk. 
\label{U2H}
\eeq
It follows from the reality of $\hs(\bx)$  that $\hhs(- \bq) = \hhs^*(\bq)$ and hence
\beq
\hbl(-\qang) = \hbl^*(\qang)\per
\label{real7}
\eeq

Eq.\ \eqref{U2H} implies that, in physical space, $\hs(\bx)$ is expressed as a convolution of $\bU(\bx)$ with a kernel $\bL(\bx)$ -- the inverse Fourier transform of $\hbl(\bq)$ -- that is homogeneous of degree $-2$, that is, $\bL(\lambda \bx) = \lambda^{-2} \bL(\bx)$.  Linear maps of this type are known as (two-dimensional) Calder\'on--Zygmund transforms \citep[e.g.][Ch.\ 2]{stein1970}. While the right-hand side of \eqref{U2H} appears ambiguous for $\bq=0$ (since $\qang$ is then not defined), we simply take $\hhs(0)=0$, corresponding to the vanishing of the spatial mean of $\hs(\bx)$, consistent with the  definition of $\hs(\bx)$  as  an anomaly.

From \eqref{linmap} and \eqref{U2H} we draw the important conclusion that patterns in $\hs$ have scales set by the scales of $\bU$ (not vorticity).  But the angular dependence of $\hs$ depends \ch{(linearly)} on the wave spectrum. In the next sections we refine this conclusion by obtaining alternative and approximate expressions for $\hbl(\bq)$.

\section{The U2H map} \label{sec:U2H} 

\subsection{Alternative forms of $\hbl(\qang)$}

A useful expression for $\hbl(\qang)$ is obtained by substituting the identity $(\bq  \bcdot \bp) \bk = (\bk \bcdot \bq) \bp - (\bk^\perp \bcdot \bp) \bq^\perp$ with $\bp=\bnabla_{\bk} \bA$ in \eqref{bL1} to obtain
\beq
\hbl(\qang) = \frac{8}{g  \ch{\bHs^2}}  \left( \int \frac{\sigma \bk \bcdot \bq}{\bc \bcdot \bq - \ii \mu} \bnabla_\bk \bA \, \dd \bk 
- \int \frac{\sigma \bk^\perp \bcdot \bnabla_\bk \bA}{\bc \bcdot \bq - \ii \mu} \, \dd \bk \, \bq^\perp
\right).
\label{bL30}
\eeq
Noting that  $\bc = \sigma \bk/(2 k^2)$,  and that the multiplication of $\mu$ by a positive factor is irrelevant, we rewrite \eqref{bL30} as
\beq
\hbl(\qang) = \frac{16}{g  \ch{\bHs^2}}  \left( \int \frac{k^2 \bk \bcdot \bq}{\bk \bcdot \bq - \ii \mu} \bnabla_\bk \bA \, \dd \bk 
- \int \frac{k^2 \bk^\perp \bcdot \bnabla_\bk \bA}{\bk \bcdot \bq - \ii \mu} \, \dd \bk \, \bq^\perp
\right).
\eeq
Now, $\mu$ can be safely set to $0$ in the first integral which, on integrating by parts, reduces to
\beq
\int k^2 \bnabla_\bk \bA \, \dd \bk =- 2 \bm{P},
\eeq
where 
\beq
\bP \defn \int \! \bA(\bk) \bk \, \dd \bk
\label{wavmom3}
\eeq 
is  the wave momentum.
For the second integral we use the polar representation \ch{\eqref{k1}}
of the SGW wavevector.
Noting that $\bnabla_\bk \bA = \partial_k \bA \,\bk/k + \partial_\theta \bA \, \bk^\perp/k^2$ and that $\bk \bcdot \bq = k q \cos(\theta-\qang)$ we obtain
\beq
\hbl(\qang) = -\frac{16}{g  \ch{\bHs^2}} \left( 2 \bm{P}  +  \int \frac{k \partial_\theta \bA}{\cos(\theta-\qang) - \ii \mu} \, \dd \bk \,  \eqp  \right)\com
\label{bL2}
\eeq
where
\beq
\eqp \defn \begin{pmatrix}-\sin \qang \\ \ \  \cos\qang\end{pmatrix}
\eeq
is the unit vector perpendicular to the wavevector $\bq$. 

Starting from \eqref{bL2} we can derive an explicit expression for the transfer function $\hbl(\qang)$ as a Fourier series in $\qang$.  We rewrite \eqref{bL2} as
\beq
\hbl(\qang) = -\frac{16}{g  \ch{\bHs^2}} \left( 2 \bm{P}  +  \,  \eqp  \, \partial_\qang \int_0^{2\pi} \frac{\Pc(\theta)}{\cos(\theta-\qang) - \ii \mu} \, \dd \theta \right) \com
\label{JV7}
\eeq  
where
\beq
\Pc(\theta) \defn \int_0^\infty\!\!\!  \bA(k,\theta) \, k^2 \dd k\per
\label{B}
\eeq
The wave momentum in \eqref{wavmom3} is
\beq
\bm{P} = \oint \Pc(\theta) \begin{pmatrix} \cos \theta \\ \sin \theta\end{pmatrix} \dd \theta\per
\eeq

With the results above  $\hbl(\qang)$ depends on the leading-order action spectrum $\bA(\bk)$ only through the function $\Pc(\theta)$. This function can be expanded in Fourier series as
\beq
\Pc(\theta) = \sum_{n=-\infty}^\infty p_n \, \ee^{n \ii \theta}, \quad \textrm{with} \quad
 2 \pi p_n = \int_0^{2\pi} \!\!\! \Pc(\theta) \, \ee^{-n \ii \theta} \, \dd \theta\per
 \label{bAn}
\eeq
Computations detailed in Appendix \ref{app:derivL3} express  the integral in \eqref{JV7} (as $\mu \to 0^+$) in terms of the $p_n$. This puts  the transfer function into the form
\beq
\hbl(\qang) =  \frac{16}{g  \ch{\bHs^2}} \left(\eqp \,   \sum_{n=-\infty}^\infty  n (-\ii)^{|n|}\,  2 \pi  p_n \, \ee^{n \ii \qang}   -  2 \bP\right)\com 
\label{bL3}
\eeq
where the wave momentum is
\beq
\bP = \begin{pmatrix} +\Re  2 \pi  p_1 \\  -\Im  2 \pi  p_1 \end{pmatrix}.
\label{Pb1}
\eeq
Equation \eqref{bL3} provides  the transfer function $\hbl(\qang) $ in a 
\ch{form readily computable from any given background wave spectrum $\bA(k,\theta)$.}

\subsection{Contributions of the divergent and vortical parts of current}\label{sec:Helm}

The 2D Helmholtz decomposition of $\bU$ into divergent and vortical parts is
\beq
\bU =  \underbrace{\bnabla \phi}_{\bU_\phi} +  \underbrace{\bnabla^{\perp} \psi}_{\bU_\psi} \com
 \label{helm0}
\eeq
where $\phi$ and $\psi$ are the potential and streamfunction, and $\bnabla^\perp = (-\p_y,\p_x)$. The corresponding decomposition of the Fourier transform $\hbU$ is
\beq
\hbU(\bq) = \ii q\hat \phi(\bq) \eq  + \ii q \hat \psi(\bq) \eqp.  \label{helm}
\eeq
In view of \eqref{linmap}, we can separate the contributions of the divergent and vortical parts of the currents by expressing the transfer function $\hbl(\bq)$ in terms of its components along $\eq$ and $\eqp$. Projecting \eqref{bL3} on $\eq$ and $\eqp$ gives
\beq
\hbl(\qang) = \hlpar \, \eq + \hlper \, \eqp, 
\label{bL2ortho}
\eeq
where
\beq
 \hlpar(\qang) =-\frac{32}{g  \ch{\bHs^2}}\bm{P}  \bcdot \eq 
\label{lparlper1}
\eeq 
and
\beq
\hlper(\qang) = \frac{16}{g  \ch{\bHs^2}} \left( \sum_{n=-\infty}^\infty n (-\ii)^{|n|} 2 \pi p_n \, \ee^{n \ii \qang} -  2 \bP \bcdot \eqp \right). 
\label{lparlper2}
\eeq
Because $\bm{e}_{-\bq} = -\eq$, the symmetry property \eqref{real7} implies that $\hlpar(-\qang)=- \hlpar^*(\qang)$ and $\hlper(-\qang)=-\hlper^*(\qang)$.

Combining the contributions proportional to $p_{\pm 1}$ (stemming from $n=\pm 1$ in the series and from $2 \bP \bcdot \eqp$), we can rewrite \eqref{lparlper2} as
\beq
\hlper(\qang) =\frac{16}{g  \ch{\bHs^2}}  \sum_{n=- \infty}^\infty n (-\ii)^{|n|} 2 \pi \tilde p_n \, \ee^{\ii n\qang},
\label{Lp17}
\eeq
where 
\beq
\tilde p_n = \begin{cases} 2 p_{\pm 1} \com \qquad &\text{if $n= \pm 1$;}\\
p_n\com \qquad   &\text{if $n \not= \pm 1$.}
\end{cases}  
\label{pntilde}
\eeq

With the form \eqref{bL2ortho} for $\hbl(\qang)$ and the Helmholtz decomposition \eqref{helm}, the linear map \eqref{linmap} becomes
\beq\label{helmhs}
\frac{\hhs(\bq)}{\ch{\bHs}} = \ii q \hlpar(\qang) \hat \phi(\bq) +  \ii q  \hlper(\qang) \hat\psi(\bq) \per
\eeq
$\hlpar$ and $\hlper$ control the dependence of $\hs$ on, respectively, the divergent and vortical parts of the current.
In general, 
$\hlpar, \, \hlper \not= 0$, and both the divergent and vortical parts of the current induce modulations in SWH. 
However, we show in \S\ref{subsec:directional} that for highly directional SGW spectra $\hlper \gg \hlpar$,  i.e.\ the vortical part of the current is dominant.

\section{Application to specific currents} \label{sec:particularflows}

Given  the wave spectrum $\bar{\mathcal{A}}(\bk)$ and the current $\bU(\bx)$, the U2H map $\hhs \ch{/\bHs} = \hbl \bcdot \hbU$,  with $\hbl$ in   \eqref{bL3} or \eqref{bL2ortho}--\eqref{lparlper2}, enables the computation of  $\hs$.  In this section, we carry out this computation. We first use a numerical procedure suitable for arbitrary currents \ch{which we apply to two realistic configurations.
We then} consider \ch{the idealised} cases of purely divergent and purely vortical currents for which we obtain analytic results. \ch{In all cases we compare the U2H predictions with the results of WW3 simulations.}

\subsection{Numerical implementation for arbitrary current} \label{sec:numimp}

The velocity field $\bU(\bx)$ is discretised on a regular grid and its Fourier transform $\hat{\bU}(\bq)$ is obtained on the dual Fourier grid by Fast Fourier transform (FFT). 
\ch{To prevent numerical artefacts due to the non-periodicity of the currents, we use a large computational domain, zero-padding $\bU$ in the periphery.}

The inverse FFT of the product $\hhs(\bq) \ch{/\bHs}= \hbl(\qang) \bcdot \hat{\bU}(\bq)$  yields  $\hs(\bx)$ on the spatial grid.
A Jupyter Notebook of this implementation is available on \url{https://shorturl.at/bef14}, where users can customize the input currents and background wave spectrum. \ch{We refer the reader to this Notebook for complete implementation details.}

\ch{For the examples of this paper}, we take the background wave action spectrum $\bA(k,\theta)$ of the separable (in $k$ and $\theta$) form detailed in appendix \ref{app:LHCS}.  The wavenumber dependence is defined by a truncated Gaussian in $\sigma(k)$ and the angular dependence $D(\theta)$ follows the model of \citet[][LHCS hereafter]{LHCS1963} 
\ch{
\beq
D(\theta) \propto \cos^{2s} \left( (\theta-\theta_p)/2\right),
\label{LHCSshort}
\eeq
}
\ch{where $\theta_p$ is the primary angle of wave propagation, measured from the $x$-axis and in the direction of $\bk$, and} the parameter $s$ controls the directional spread. Large values, say $s \gtrsim 10$, correspond to swell-like sea states. \ch{For integer $s$, the coefficients $p_n$ in \eqref{bAn} required for $\hbl(\qang)$ have a simple closed form and vanish for  $|n|>s$  (see appendix \ref{app:LHCS}).}

\ch{
For comparison with U2H, we carry out WW3 simulations that approximate a steady solution of the linear action equation \eqref{actioncons}. The set up is as described in \citet{WVBYV} except for two aspects of the wave forcing. First,} the forcing imposes the spectrum $\bar{\mathcal{A}}(\bk)$ on the entire boundary i.e. waves enter the rectangular domain from all four sides. This improved formulation  ensures that the wave spectrum in the absence of currents is  uniform even for spectra with broad directional spread. (In \citet{WVBYV} waves entered the computational domain only from the western boundary. Even in the complete absence of currents the resulting steady-state solution decreased with $x$ as `wave shadows'  from the northern and southern boundaries encroach into the centre of the domain.)  \ch{Second, to ensure 
consistency with U2H, we zero-pad the domain of the currents in strips with widths of 4 grid spacings so that waves are forced at current-free boundaries. For both U2H and WW3 we report SWH anomalies obtained by subtracting the spatial average over the (unpadded)  domain shown in the figures.}

\ch{We apply the U2H and the WW3 implementations on two examples of realistic currents. The first example,  already shown in figure \ref{fig:hs_mitgcm}, uses a snapshot of currents in the California Current system simulated from MITgcm, as configured in  \citet{VillasBoas2020}.}
The current speeds reach $0.65$ m s$^{-1}$ at maximum.  The waves are forced with with peak period $10.3$ s corresponding to a  wavelength of 166 m and a group speed of 8 m s$^{-1}$. \ch{The waves are swell-like with parameter $s=10$, and propagate primarily in the direction $\theta_p=0$. The U2H prediction of $\hs$ (panel c) is in good agreements with that from WW3 (panel b), with difference field (panel d) lower than $7\%$ in amplitude of $\hs/\bHs$. The  SWH anomalies predicted by U2H have larger overall amplitudes than of WW3. This  is reflected in the probability densities shown in  figure \ref{fig:hspdf} (left panel), which confirm that U2H predicts more extreme values than WW3. We tentatively attribute this to numerical damping effects from WW3.}

\begin{figure}
  \centering
  \includegraphics[width=1.0\textwidth]{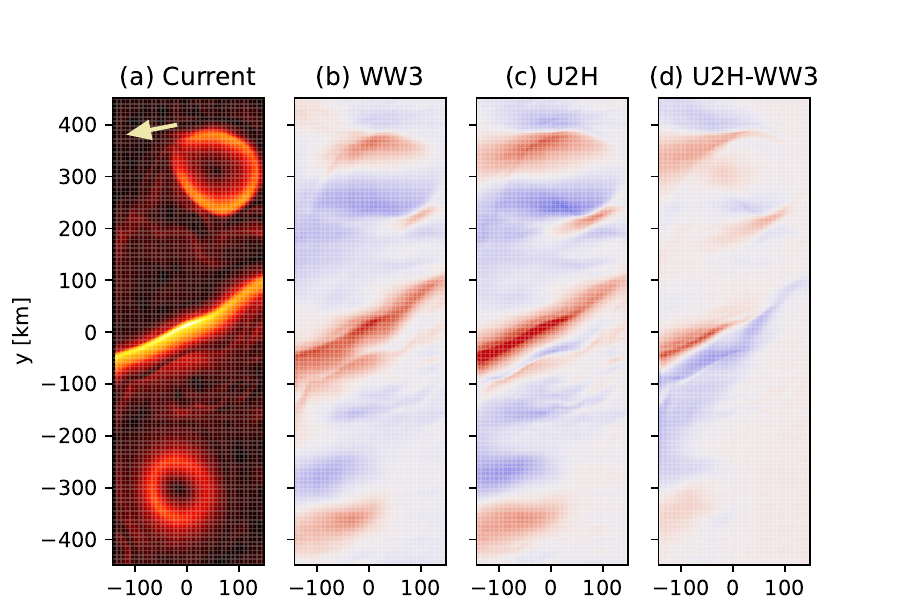}
  \caption{\ch{(a): Surface current speeds in an MITgcm  simulation of the Gulf Stream,  with the arrow indicating  the primary  direction of  wave propagation. SWH anomaly computed using (b) WW3  and (c) the U2H map. (d): difference between (c) and (b). 
The background waves use the LHCS model spectrum \eqref{LHCS111} with $s=16$ and peak angle $\theta_p=191 ^{\circ}$.}}
  \label{fig:hsgulfstream}
\end{figure}

\begin{figure}
  \centering
  \includegraphics[width=0.45\textwidth]{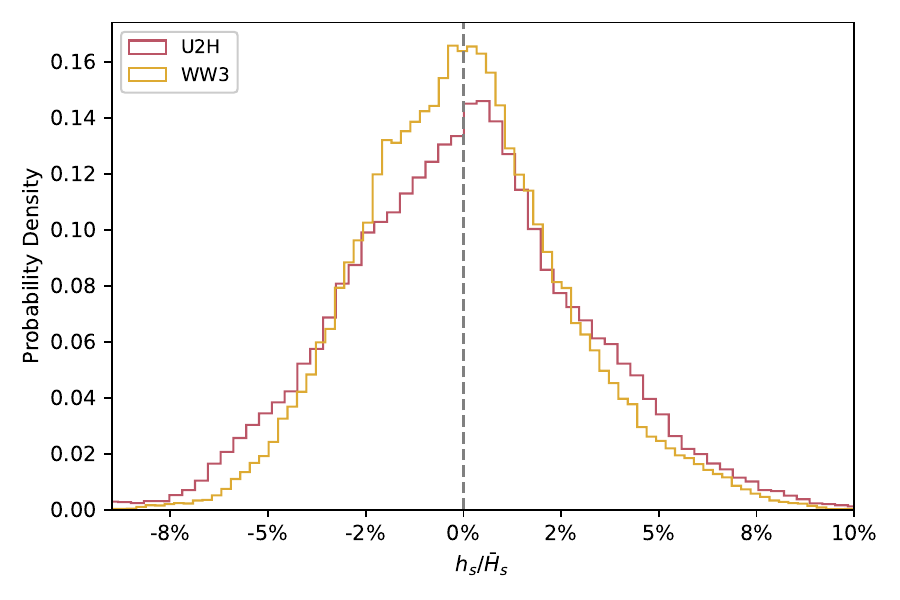}
    \includegraphics[width=0.45\textwidth]{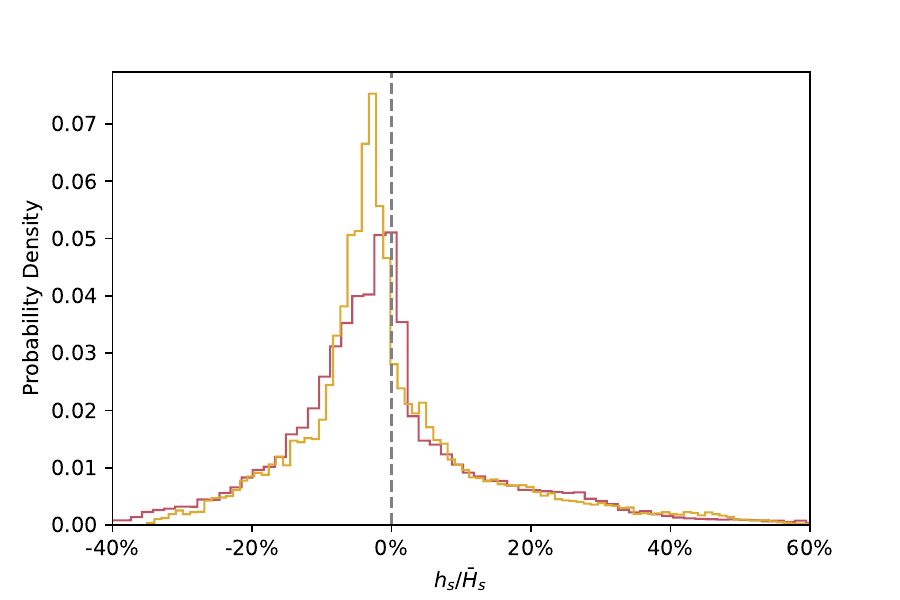}
    \caption{\ch{Estimated probability densities for $\hs$ computed using the U2H map (red lines) and WW3 model (yellow lines), for the example shown in figure \ref{fig:hs_mitgcm} (left panel) and figure \ref{fig:hsgulfstream} (right panel).  Probability densities are estimated by grouping  the values of $\hs/\bHs$ within the unpadded domains into 100 bins.}
    }
  \label{fig:hspdf}
\end{figure}

\ch{The second example, shown in figure \ref{fig:hsgulfstream}, uses a Gulf Stream current's snapshot from the MITgcm simulation in figure 1 of \citet{Ardhuin2017}. The current  speeds reach $2.9$ m s$^{-1}$ at maximum. The waves are forced with peak period of  $14.3$ s corresponding to a wavelength of $319$ m and group speed of $11$  m s$^{-1}$, and are swell-like, with  parameter $s=16$ and  $\theta_p=191 ^{\circ}$. These parameters are estimated from buoy data for the same time as for figure 1 in \citet{Ardhuin2017}.
Although we use a similar current  snapshot and wave forcing as  \citet{Ardhuin2017}, our WW3 configuration is different from theirs, and disagreements in $\hs$ are expected.} 
\ch{In this example, the SWH anomalies are large, with $\hs/\bHs$ exceeding 50\% in some locations, challenging our assumption of linearity. Nonetheless, there is a good qualitative match between the WW3 and U2H results. The largest differences arise in regions of high-speed currents. The probability density functions from the U2H and WW3 outcomes (figure \ref{fig:hspdf}, right panel) are skewed differently.  These differences may be attributed to higher order terms neglected by U2H.}

We now consider idealised scenarios to gain insight into the dependence of $\hs$ on $\bU$.

\subsection{Divergent  current} \label{sec:irrot}

For a purely divergent current, with $\bU_{\psi}=0$, \eqref{helmhs} reduces to 
\beq
\hhs(\bq) = -\frac{32}{g  \bHs}  \ii q   \hat \phi(\bq)\, \bm{P}  \bcdot \eq  \per
\label{irrot7}
\eeq	
 Since $\ii q  \eq  = \ii \bq$,  the inverse Fourier transform of \eqref{irrot7} is
\beq
\hs =  -\frac{32}{g  \bHs} \bU_{\phi}\bcdot \bP \per
\label{irrot11}
\eeq
Thus the SWH anomaly that arises in response to a divergent current is proportional to the component of  current velocity along the wave momentum. In particular, the response is local and vanishes where the current vanishes.


\begin{figure}
  \centering
  \includegraphics[width=1.0\textwidth]{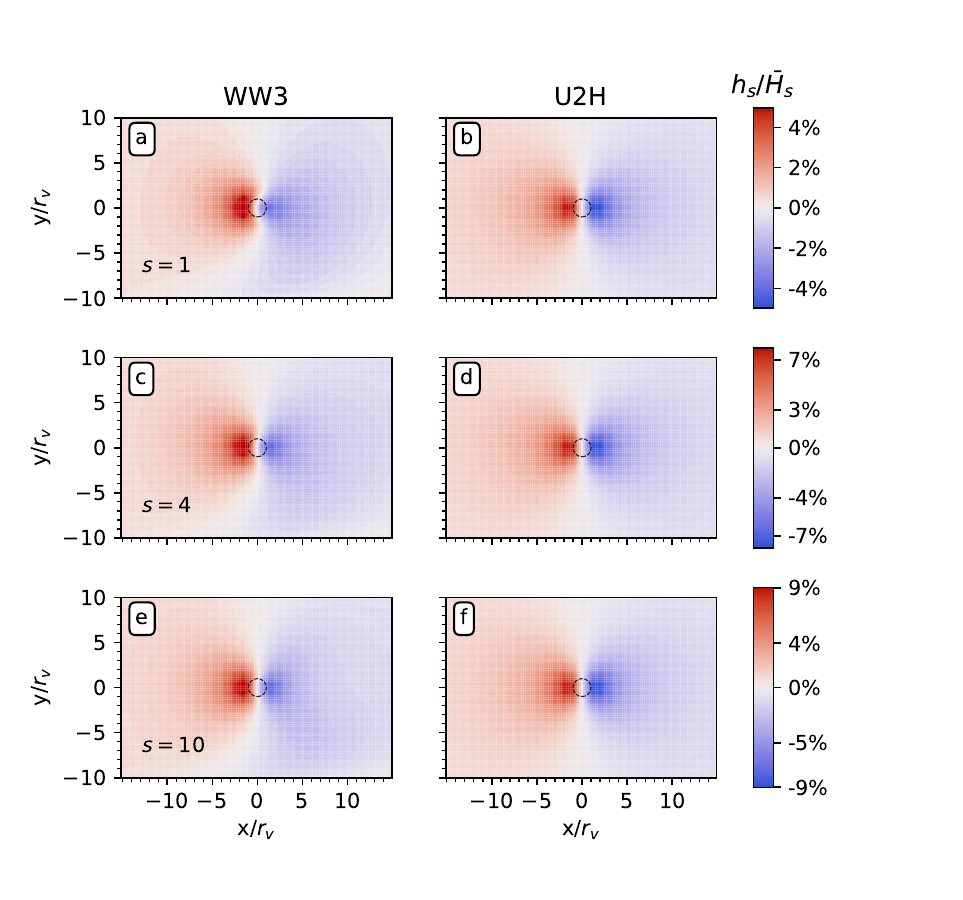}
    \caption{SWH anomaly for the divergent flow with Gaussian divergence \eqref{gaussdiv} with characteristic radius $r_v = 25$ km (indicated by the dashed circle) and maximum speed $0.8$ m s$^{-1}$. The results of WW3 simulations (left) are compared with the U2H prediction \eqref{irrot11} (right) for three values of the parameter $s$ characterising the \ch{directional width} of the wave spectrum.}
  \label{fig:hs_GD}
\end{figure}

We illustrate \eqref{irrot11} with a simple axisymmetric, divergent current whose divergence is the Gaussian
\beq
\bnabla \bcdot \bU = \nabla^2 \phi =\frac{\kappa}{2\pi \rv^2} \, \ee^{-r^2/2\rv^2},
\label{gaussdiv}
\eeq
where $\rv=25\, \text{km}$ is the characteristic radius and $\kappa$ is the {area flux}, set such that the maximum current speed $U_m=\sqrt{U^2+V^2}$ is $0.8$ m s$^{-1}$.

Figure \ref{fig:hs_GD} compares the U2H prediction \eqref{irrot11} for this current with results for WW3 simulations for three values of the directionality parameter $s$. Figure \ref{fig:hs_GD}  confirms the validity of the U2H prediction and the local nature of the SWH response. This response  to divergent currents   has a spatial structure independent  of the directional  spread of wave energy i.e. $\hs$  in \eqref{irrot11} depends only on $\bP$. This   striking result is in sharp contrast with the response to  vortical currents as we show next.

\subsection{Vortical current}\label{sec:GV}

For a purely  vortical currents, $\bU_{\phi}=0$ in \eqref{helm0} and the U2H map in \eqref{helmhs} is determined by the scalar transfer function $\hlper(\varphi)$, which is explicitly computed from  the series in \eqref{Lp17}. 
As a demonstration, we consider a Gaussian vortex, with zero divergence and vorticity in physical and Fourier space given by
\beq \label{Gzeta}
\zeta(\bx)=\frac{\kappa}{2\pi r_{\mathit{v}}^2} \ee^{-r^2/(2r_{\mathit{v}}^2)}
\quad
\textrm{and}
\quad
\hat \zeta (\bq) = \kappa \ee^{-r_{\mathit{v}}^2 q^2/2},
\eeq
where $\kappa$ is the circulation. 
We take advantage of the axisymmetry of this flow to carry out the Fourier inversion leading to $\hs(\bx)$ analytically. Calculations detailed in Appendix \ref{app:GVhs} yield the explicit expression
\beq
\hs(\bx)=-\frac{16 \ii}{g \bHs } \frac{\kappa}{\rv} \sqrt{\frac{\pi}{2}} \ee^{-r^2/4 r_v^2}   \sum_{n=-\infty}^\infty  n \,  \tilde p_n  \,   I_{|n|/2}(r^2/4 r_v^2)     \, \ee^{\ii n \xang},
\label{Bess7}
\eeq
where $\bx = r (\cos \nu,\sin \nu)$ and the $I_{|n|}$ are modified Bessel functions. The coefficients $\tilde p_n$ depend only on the wave spectrum and are related to the already obtained $p_n$ according to \eqref{pntilde}. Formula \eqref{Bess7} has the advantage over the general implementation of the U2H map described in \S\ref{sec:numimp} in that it gives  $\hs(\bx)$ at any location without the need for entire computational domains in both physical and Fourier domains.

\begin{figure}
  \centering
  \includegraphics[width=1.0\textwidth]{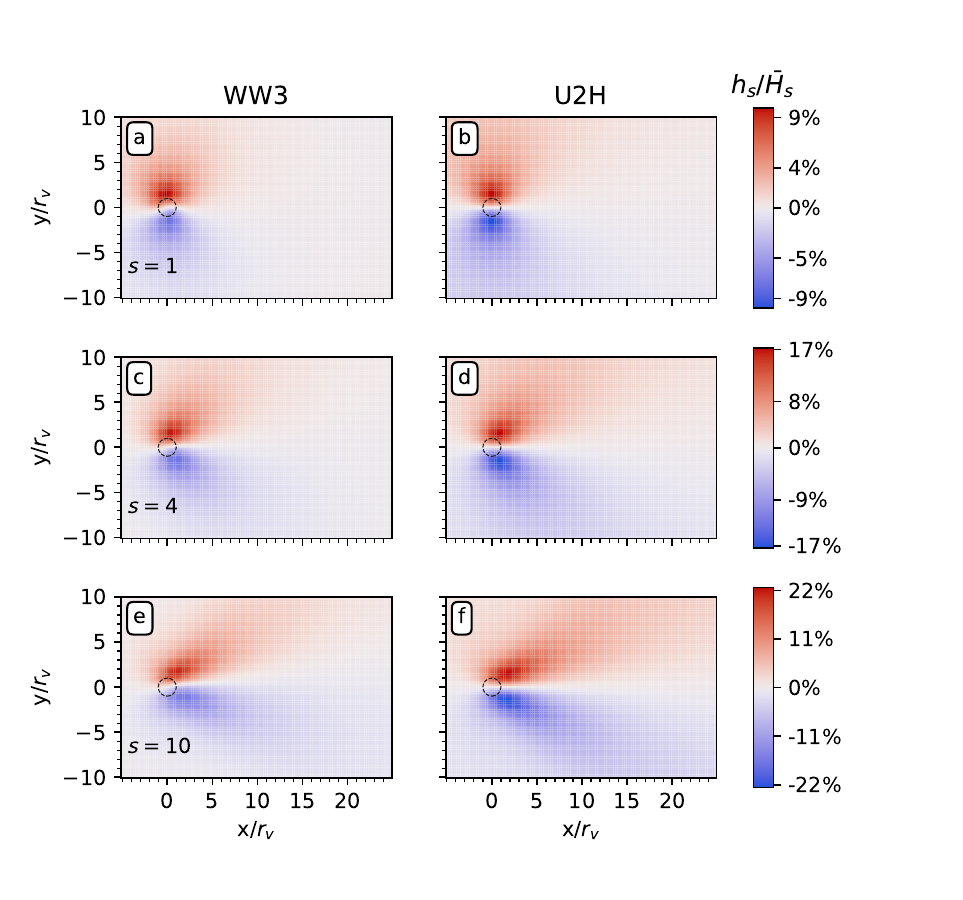}
    \caption{Same as figure \ref{fig:hs_GD} but for the  Gaussian vortex with $\zeta(r)$ in  \eqref{Gzeta}.}
  \label{fig:hs_GV}
\end{figure}

Figure \ref{fig:hs_GV} compares the SWH anomaly  \eqref{Bess7} with that obtained in WW3 simulations. The parameters $r_v = 25$ km and $U_m = 0.8$ m s$^{-1}$ are the same as those of the divergent flow in \S\ref{sec:irrot}. The SWH response in figure \ref{fig:hs_GV}    is very different from the response to divergent currents in figure  \ref{fig:hs_GD}.  The SWH anomaly in figure \ref{fig:hs_GV} extends beyond the vortex. For the swell-like case $s=10$ there is  a wake-like feature decaying  slowly in the direction of wave propagation. This physically important limiting case is discussed in \S\ref{subsec:directional} and in \citet{WVBYV}.

\section{Particular wave spectra} \label{sec:particularspectra}
In this section we examine the role of the wave spectrum $\bar{\mathcal{A}}(\bk)$ in shaping the SWH anomaly by considering special and limiting cases.

\subsection{Isotropic wave spectrum}

\begin{figure}
  \centering
  \includegraphics[width=0.45\textwidth]{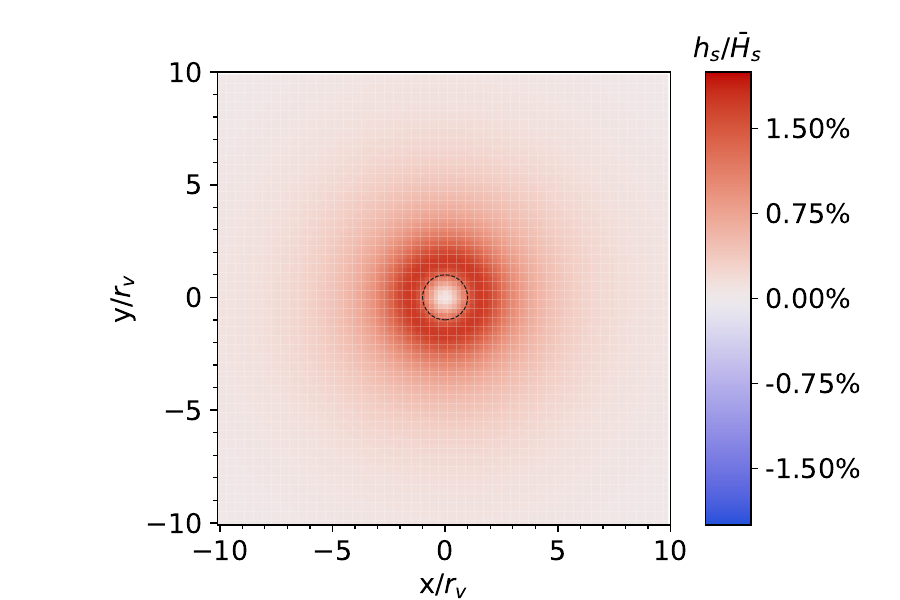}
    \includegraphics[width=0.5\textwidth]{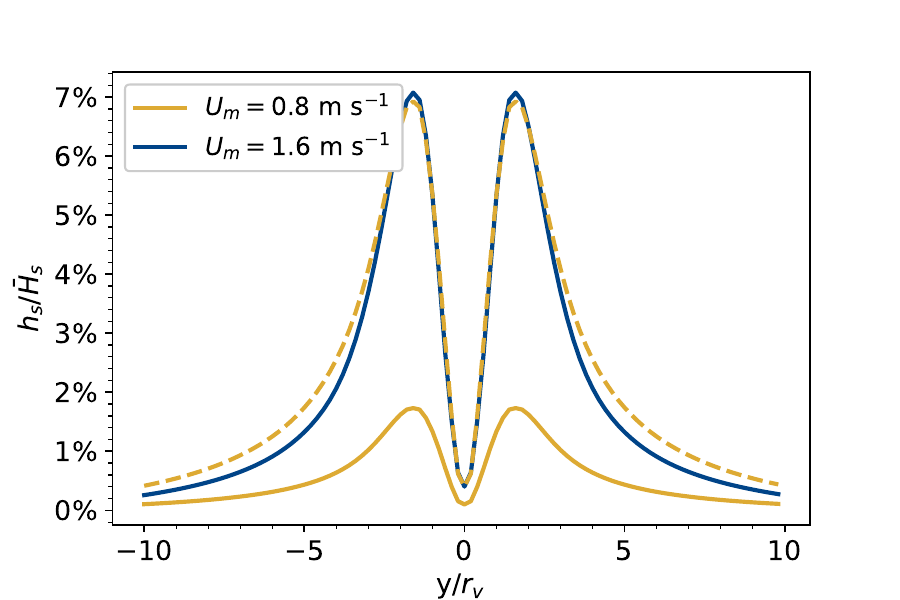}
    \caption{SWH anomaly computed using WW3 for an isotropic wave spectrum ($s=0$ in the LHCS model)  with the Gaussian vortex $\zeta(r)$ in \eqref{Gzeta}. Left panel: $\hs$ for $U_m=0.8$ m s$^{-1}$, with the dashed circle indicating the vortex radius $\rv$. Right panel: cross section of $\hs$ at $x=0$ (slicing through the centre of the  vortex) for $U_m=1.6$ m s$^{-1}$ (blue solid curve) and  $U_m=0.8$ m s$^{-1}$ (yellow solid curve). The yellow dashed curve is obtained by multiplying $\hs$  for  $U_m=0.8$~m~s$^{-1}$ by 4.}
  \label{fig:hs_iso}
\end{figure}


Eq.\  \eqref{bL2}  shows that $\hbl(\qang)=0$ if the SGW action (or energy) spectrum is isotropic since $\partial_\theta \bar{\Ac}(\bk)= 0$ and  $\bP=0$ as a consequence. Thus,  if the wave spectrum is isotropic, currents do not induce modulations of the SWHs (at the order we consider). 

To verify this, we ran WW3 simulations with the isotropic wave spectrum obtained by setting $s=0$ in the LHCS model of Appendix \ref{app:LHCS}
and the currents from either the MITgcm simulation in  figure \ref{fig:hs_mitgcm} or the Gaussian vortex of \ref{fig:hs_GV}.  The SWH anomaly $\hs$ in both cases  is at most 3\%, much smaller than found for anisotropic spectra. The small but non-zero $\hs$ for isotropic spectra\ch{is the result of effects quadratic in $\bU$. This $\varepsilon^2$-term is} not captured by the linear U2H map. 
We confirmed this by  increasing the velocity of the Gaussian vortex by a factor of 2 (setting $U_m= 1.6$ m s$^{-1}$ instead of 0.8 m s$^{-1}$) so that   $\hs$ increases by a factor 4: see  Figure \ref{fig:hs_iso} \ch(right panel). (The MITgcm outcome is not shown.)

\subsection{Mildly anisotropic wave spectrum} \label{sec:mildly}

The U2H map is particularly simple for the spectrum 
\beq
\bar{\Ac}(\bk) = \Ac_0(k) + \Ac_c(k) \cos \tha,
\label{simpdir}
\eeq
e.g., as in the LHCS spectrum with \ch{$s=1$} used for the first row of figures \ref{fig:hs_GD} and \ref{fig:hs_GV}.
For the action  spectrum in \eqref{simpdir} the wave momentum \eqref{wavmom3} can be written as
\beq
\bP = |\bP| \begin{pmatrix} 1 \\ 0 \end{pmatrix}=  |\bP| \cos \qang \,  \eq -  |\bP| \sin \qang \, \eqp,
\eeq
where
\beq
| \bP | = \pi \int  \!\! \Ac_c(k)  k^2 \dd k.
\eeq
The  function $\Pc(\tha)$ defined in  \eqref{B} is then
\beq
\Pc(\tha) = \int \!\! \Ac_0(k)  k^2 \dd k +  \frac{|\bP|}{\pi} \cos \theta.
\eeq
and $2 \pi p_1 = 2 \pi p_{-1} = |\bP|$.
 The transfer function in \eqref{bL3} reduces to 
\beq
\hbl(\qang) =   \frac{32 }{g  \ch{\bHs^2}} |\bP| \left(2 \sin \qang \, \eqp - \cos \qang \,  \eq \right)\per
\eeq
With the Helmholtz decomposition \eqref{helm}, the U2H map is
\begin{align}
\frac{\hhs(\bq)}{\ch{\bHs}}&= \hbl(\bq) \bcdot \hbU(\bq)\com \\
&= \frac{32 }{g \ch{\bHs^2}}\left(2|\bP|\,  \ii q \sin \qang\,  \hat \psi - |\bP|\,  \ii q\cos \qang\, \hat \phi \right)\per \label{eq57}
\end{align}
The inverse Fourier transform can be taken by inspection and the result written as
\beq
\hs = -\frac{32}{g\bHs} \left(2 \bU_\psi \bcdot \bP + \bU_\phi \bcdot \bP \right)\per
\label{secsimp11}
\eeq

In this simple case, only  the component of current along $\bP$ produces a SWH anomaly which turns out to be local, vanishing where the current vanishes. In \eqref{secsimp11} the vortical part of the current, $\bU_\psi$, is twice as effective as the divergent part, $\bU_\phi$. The divergent contribution in \eqref{secsimp11}  is identical to that  in \eqref{irrot11} which applies to arbitrary wave spectra.  This example, which corresponds to the first rows of  figures \ref{fig:hs_GD} and  \ref{fig:hs_GV}, shows that the response to divergent currents is not always negligible relative to the vortical response. This is in contrast with \citet{VillasBoas2020}'s suggestion that only the vortical part of the current affects $\hs$. The next section however shows that the imprint of the vortical part of the current is much larger than that of the divergent part for highly directional spectra.

\subsection{Highly directional wave spectrum}\label{subsec:directional}

We conclude above that $\hs$ is small for an isotropic wave spectrum. It is of interest to examine the opposite limit of a highly directional wave spectrum. This limit corresponds to an action spectrum of the form
\beq
\bA(k,\theta) = \delta^{-1} \bA(k,\Theta), \quad \textrm{where} \quad \Theta = \theta/\delta,
\label{direc}
\eeq
with $\delta \ll 1 $ the relevant small parameter, and we assume that $\theta=0$ is the primary propagation direction. The prefactor $\delta^{-1}$  ensures that the action spectrum integrated over $\theta$ is $O(1)$.  Correspondingly,
\beq
\Pc(\theta) = \delta^{-1} \Pc(\Theta).
\label{Basy}
\eeq
For simplicity, we abuse notation by using the same symbols $\bA$ and $\Pc$ on both sides of \eqref{direc}--\eqref{Basy}, distinguishing them by their arguments. \ch{Similar to the treatment of $\eps$ in \eqref{freq7}, we use $\delta$ as a bookkeeping parameter that is set to $1$ in the end.}

Taking \eqref{JV7} as a starting point, we obtain an asymptotic approximation to $\hbl(\qang)$ in appendix \ref{app:swell}. There we show that the dominant contribution to $\hbl(\qang)$ comes from the integral term and is large in small regions around $\qang = \pm \pi/2$. In terms of the rescaled variable
\beq
 \Phi_\pm = (\qang \mp \pi/2)/\delta=O(1),
 \label{Phiphi}
\eeq
the leading-order approximation to the transfer function in these regions is 
\beq
\hbl(\qang) \sim \hlper(\qang) \begin{pmatrix} \mp 1 \\  0 \end{pmatrix} \sim   \frac{16}{g  \ch{\bHs^2} \delta^2}  \,  \partial_{\Phi_\pm} \int_{-\infty}^\infty \frac{\Pc(\Theta)}{\Theta - \Phi_\pm \mp \ii \mu} \, \dd \Theta  \,  \begin{pmatrix}1 \\  0 \end{pmatrix}.
\label{JV8}
\eeq
Thus $\hbl(\qang)$ is dominant and $O(\delta^{-2})$ in narrow, $O(\delta)$, sectors around $\qang = \pm \pi/2$. We conclude that: 
\begin{enumerate}
\item \ For typical $\hbU(\bq)$,  patterns of $\hs$ take the form of structures elongated in the direction of propagation of the waves,  i.e.\ streaks, with an $O(\delta)$ aspect ratio.
\item \ $\hlper = O(\delta^{-2}) \hlpar$, since $\hlpar = O(1)$ (see \eqref{asybL}). According to \eqref{helmhs}, this implies that the divergent-free, vortical part of the velocity field $\bU$ has an asymptotically larger impact on $\hs$ than the potential part.
\item \ Highly directional waves   produce   SWH anomalies larger by  a factor $\delta^{-1}$  than those induced for spectra with $O(1)$ directional spread. (This estimate accounts for both the factor $\delta^{-2}$ in \eqref{asybL}  and the $O(\delta)$ width of the support of $\hlper(\qang)$ implied by \eqref{Phiphi}). 
\item \ \ch{As a result of (iii), the linear approximation that underpins U2H requires that $\eps \ll \delta$ in addition to $\eps \ll 1$. We discuss this further at the end of the section.}
\end{enumerate}


We illustrate the asymptotic result \eqref{asybL} by considering the limit $s \to \infty$ of the LHCS spectrum. Taking this limit in \eqref{Pthetas} gives
\beq
\mathcal{P}(\Theta) = \frac{\alpha}{\sqrt{2\pi}} \ee^{-\Theta^2/2}, \quad \textrm{where} \quad \delta =\sqrt{2/s}
\label{Pswell}
\eeq
and $\alpha$ a constant determined by the dependence of the spectrum on $k$. 
Using \eqref{Pswell} and  the Sokhotski–Plemelj theorem, we rewrite the integral term in \eqref{asybL} as
\begin{align}
 \int_{-\infty}^\infty \frac{\Pc(\Theta)}{\Theta - \Phi_\pm \mp \ii \mu} \, \dd \Theta 
 &= \frac{\alpha}{\sqrt{2 \pi}} \left( \pm \ii \pi \, \ee^{-\Phi_{\pm}^2/2} + \dashint_{-\infty}^\infty \frac{\ee^{-\Theta^2/2}}{\Theta-\Phi_{\pm}}\,\dd\Theta \right) \nonumber \\
 &= \alpha \left( \pm \ii \sqrt{\pi/2} \, \ee^{-\Phi_{\pm}^2/2} - \sqrt{2} \, \mathrm{daw}(\Phi_{\pm}/\sqrt{2}) \right),
 \label{plemelj}
\end{align}
where $\dashint$ denotes the Cauchy principal value and $\mathrm{daw}(\cdot)$ denotes the Dawson function \citep{DLMF}. Using that $\mathrm{daw}'(x)=1-2x \, \mathrm{daw}(x)$ \citep{DLMF} we can evaluate the right-hand side of \eqref{asybL} to find 
\beq
\hlper(\varphi) = \frac{16 \alpha}{g  \ch{\bHs^2}\delta^2} \begin{cases}
\ii \sqrt{\pi/2} \, \Phi_{+} \ee^{-\Phi_{+}^2/2} + 1 - \sqrt{2}\Phi_{+} \, \mathrm{daw}(\Phi_{+}/\sqrt{2}),  &\textrm{for}\ \ 0\leq\qang<\pi;\\
\ii \sqrt{\pi/2} \, \Phi_{-} \ee^{-\Phi_{-}^2/2} - 1+ \sqrt{2}\Phi_{-} \, \mathrm{daw}(\Phi_{-}/\sqrt{2}), &\textrm{for} \ \ -\pi<\qang<0 \per
\end{cases}
\label{LswellLHCS}
\eeq

\begin{figure}
  \centering
  \includegraphics[width=1.0\textwidth]{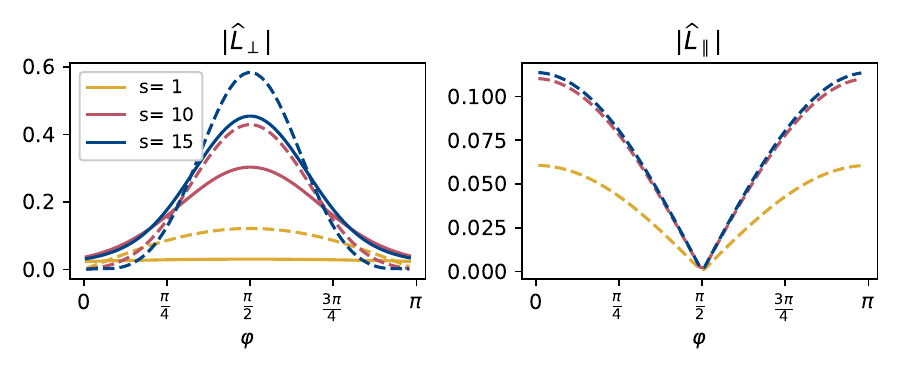}
    \caption{Magnitudes of the transfer functions $\hlper(\qang)$ (left) and $\hlpar(\qang)$ (right) associated with the vortical and potential part of the current as functions of $\qang$ for the LHCS spectrum with directionality parameter $s=1, \, 10$ and $15$. The exact values computed from  \eqref{lparlper1} and \eqref{lparlper2} are shown by the dashed lines; the solid lines in the left panel show the large-$s$ approximation \eqref{LswellLHCS} for $\hlper(\qang)$. We take advantage of the symmetry \eqref{real7} to show only the range $\qang \in [0,\pi]$. 
    }
  \label{fig:Lasymp}
\end{figure}

Figure \ref{fig:Lasymp} compares the asymptotic approximation \eqref{LswellLHCS} of $\hlper$ with the exact values obtained from \eqref{lparlper2} for $s=1$, $10$ and $15$. It shows the asymptotic approximation to be reasonably accurate for $s = 10$. We have checked that the error scales as $O(\delta^2)$. The figure also shows $\hlpar$ in \eqref{lparlper1} to confirm that $\hlper \gg \hlpar$, and hence that vortical part of the current dominates over the divergent part, for $s \gg 1$. 

\begin{figure}
  \centering
  \includegraphics[width=1.0\textwidth]{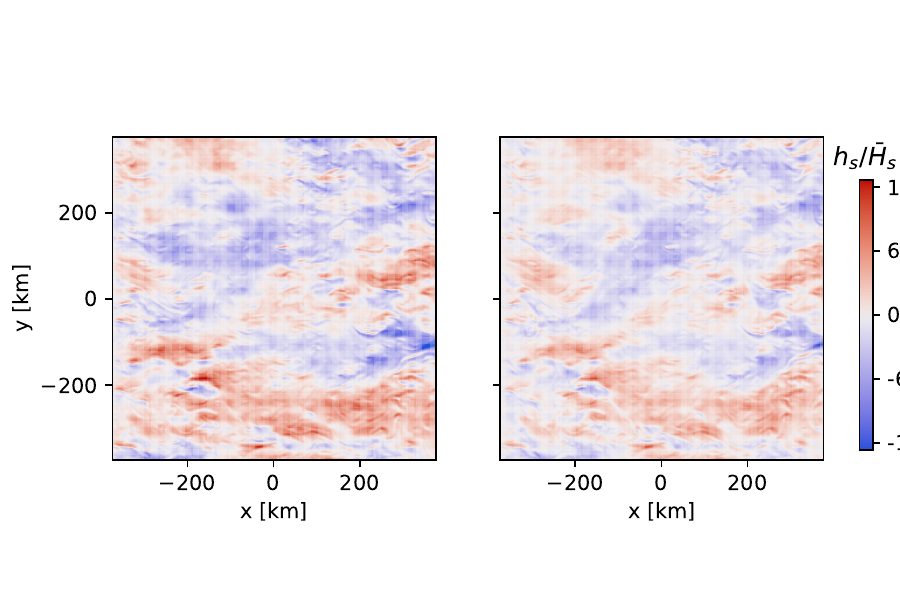}
    \caption{SWH anomaly for \ch{a highly directional} ($s=10$) wave spectrum and MITgcm current of figure \ref{fig:hs_mitgcm} computed with the full U2H map  (left panel, identical to panel (c) of figure \ref{fig:hs_mitgcm}), and with the $s \gg 1$ asymptotic approximation (right panel). This figure can be produced from the notebook  accessible at \url{https://shorturl.at/bef14}.}
  \label{fig:hs_mitgcm_swell}
\end{figure}
 
As an application of \eqref{LswellLHCS}, in figure \ref{fig:hs_mitgcm_swell} we compare the predictions of the U2H map for the MITgcm simulation current of figure \ref{fig:hs_mitgcm} computed with the exact $\hbl$ 
and with the asymptotic approximation \eqref{LswellLHCS}. The match is very good, even though for $s=10$, $\delta \approx 0.45$ is only marginally small. The code applying the expression \eqref{LswellLHCS} is available on the Jupyter Notebook \url{https://shorturl.at/bef14}, where readers can also experiment with different choices of the parameter $s$ to observe how the agreements get better/worse with larger/small $s$.

We conclude by connecting the results of this section with those of \citet{WVBYV}. They focus on the regime $\delta \ll 1$ and on localised currents. Using matched asymptotics, they obtain an asymptotic expression for the total SWH $\Hs = \bHs + \hs$ in the presence of currents. This expression holds without the linearity assumption $\hs \ll \bHs$ that underpins the U2H map. Specifically, they consider the distinguished limit $\delta = O(\eps)$ which leads to $\hs/\bHs = O(1)$. \citet{WVBYV} give a simplified form valid when $\eps \ll \delta \ll 1$. In appendix \ref{app:swell} we show that U2H in the approximation \eqref{asybL} reduces to this form for localised currents.

\section{Discussion and conclusion}

In the oceanographic regime with $U/\ch{c_g} \ll 1$ the effect of currents  on an underlying spatially uniform action spectrum $\bar \Ac(\bk)$ can be determined by solving the linear problem in \eqref{a_t} and \eqref{a_t2} for the anomaly in action density $a(\bx,\bk)$. We have  focussed on extraction of the anomaly in SWH $\hs(\bx)$ via the weighted $\bk$-integral of $a(\bx,\bk)$ in \eqref{hsa}. The results are in  good agreement with numerical solutions of WW3. There is a  significant generalization of this procedure:  given $a(\bx,\bk)$, other important SGW  properties, such as the current-induced anomaly in the Stokes \ch{drift},  are only a  $\bk$-integral away.

\ch{Assumptions involved in the U2H map, listed in \S \ref{sec:formulation}, are violated in \ch{several} ocean-relevant scenarios: \ch{interactions of surface waves with} tidal and near-inertial currents violate the assumption of steady current \citep[e.g.,][]{Tolman1988, Tolman1990, Ho2023, halsne2024wave,gemmrich2012signature}; short surface waves in the saturation range of the wave spectrum violate the assumptions of scale separation and large group speed \citep[e.g.,][]{rascle2017intense, lenain2021modulation, vrecica2022observations}; and conditions of active wave generation and dissipation make the source terms non-negligible \citep[e.g.,][]{holthuijsen1991effects, chen2007cblast, Romero2017}. These scenarios are not explored in this work. Nevertheless, for relatively long surface waves interacting with open-ocean currents, our assumptions are often satisfied.}

Linearity of the   partial differential equation  \eqref{a_t} implies that there is a linear  map from $\bU$ to  $\hs$. Can this U2H map be inverted to produce an H2U map?
Not in general: U2H maps a vector field to a scalar field, so cannot be expected to have an inverse.
The non-invertibility of U2H  is illustrated by considering the mildly anisotropic spectrum of \S\ref{sec:mildly}:
\eqref{eq57} implies that $\hs=0$ for a velocity field with potential and streamfunction satisfying $ 2 \psi_y - \phi_x = 0$, demonstrating the non-uniqueness of $\bU$ for a given $\hs$.  Observations of $\hs$, however, provide partial information about $\bU$. For swell-like waves, in particular, \S \ref{subsec:directional} shows that $\hs$ can be approximated by a linear operator acting on the vortical component of the current; in this case, we can infer vorticity from $\hs$.

An  important qualitative  result emerging  quickly from the analysis is that  the transfer function, $\hbl$ in \eqref{U2H},   does not depend on the magnitude $q$ of the current wavenumber $\bq$, but only on its direction $\varphi$. This implies that the spatial scale of variations in  $\hs$ is set by those of the current $\bU$, e.g.\ power-law $\bU$-spectra result in power-law $\hs$-spectra with the same slope \citep{Ardhuin2017, Romero2020, VillasBoas2020}. Using \eqref{helmhs} we are now exploring the ramifications of this result.

\backsection[Acknowledgements] {\ch{The authors
thank Fabrice Ardhuin for providing the Gulf Stream current field used in figure 2. Comments from the reviewers have substantially improved the manuscript. }}

\backsection[Funding]{JV and HW are  supported by the UK Natural Environment Research Council (grant NE/W002876/1). ABVB is supported by NASA award 80NSSC23K0979 through the International Ocean Vector Winds Science Team and NASA award 80NSSC24K0411 through the S-MODE Science Team. WRY is supported by the National Science Foundation award 2048583. }

\backsection[Declaration of interests]{The authors report no conflict of interest.}

\backsection[Data availability statement]{The customizable Jupyter Notebook that computes the U2H map  from $\bU$ to $\hs$ under an LHC incoming wave spectrum is available on \url{https://shorturl.at/qrxEJ}. The configuration files used to produce the outputs from WW3 used in this work are available at \url{https://github.com/biavillasboas/U2H}.}

\backsection[Author ORCID]{H. Wang, https://orcid.org/0000-0002-5841-5474. A. B. Villas B\^{o}as, https://orcid.org/0000-0001-6767-6556.  W.R Young,  https://orcid.org/0000-0002-1842-3197.  J. Vanneste, https://orcid.org/0000-0002-0319-589X.}

\appendix

\section{Computable form of the transfer function $\hbl$} \label{app:derivL3}

Introducing \eqref{bAn} into the integral in \eqref{JV7} and changing integration variable from $\theta$ to $\theta + \qang$ gives 
\beq\label{limint}
\int_0^{2\pi} \frac{\Pc(\theta)}{\cos(\theta-\qang) - \ii \mu} \, \dd \theta  =\sum_{n=-\infty}^\infty  \left(  \int_0^{2\pi} \frac{\ee^{n \ii \theta}}{ \cos \theta - \ii \mu} \, \dd \theta \right) \, p_n \, \ee^{n\ii \qang}.
\eeq
The integrals on the right-hand side can be evaluated explicitly using contour integration in the complex plane. Consider first $n\ge0$ and let $z=\ee^{\ii \theta}$ so that
\beq
\int_0^{2\pi} \frac{\ee^{n \ii \theta}}{ \cos\theta - \ii \mu} \, \dd \theta = - 2 \ii \oint \frac{z^n}{z^2-2\ii \mu z +1} \, \dd z.
\label{contour}
\eeq
We can now use  the residue theorem, noting that the integrand has to the two poles $z_\pm = \pm \ii + \ii \mu + O(\mu^2)$ and that only $z_-$ is enclosed by the integration contour since $\mu \to 0^+$. This leads to
\beq
\lim_{\mu \to 0^+} \int_0^{2\pi} \frac{\ee^{n \ii \theta}}{ \cos \theta - \ii \mu} \, \dd \theta = 2 \pi (-\ii)^{n-1} \quad \textrm{for} \ \ n \ge 0.
\label{cauchy}
\eeq
For $n<0$, we can let $\theta \mapsto - \theta$ on the left-hand side of \eqref{contour}  to conclude that \eqref{cauchy} holds with $n \mapsto -n$. Hence, for any $n$,
\beq
\lim_{\mu \to 0^+} \int_0^{2\pi} \frac{\ee^{n \ii \theta}}{ \cos \theta - \ii \mu} \, \dd \theta = 2 \pi (-\ii)^{|n|-1},
\label{cauchy2}
\eeq
and \eqref{limint} reduces in the limit $\mu \to 0^+$ to
\beq
\lim_{\mu \to 0^+}  \int_0^{2\pi} \frac{\Pc(\theta)}{\cos(\theta-\qang) - \ii \mu} \, \dd \theta  =
  2 \pi \sum_{n=-\infty}^\infty (-\ii)^{|n|-1} p_n \, \ee^{n\ii \qang}.
\label{aa}
\eeq
Taking the derivative in $\varphi$ lead to the coefficient in front of $\eqp$ in \eqref{bL3}.

The computation of the wave momentum $\bm{P}$ in terms of the Fourier coefficients of the action spectrum is straightforward: using \eqref{wavmom3}, \eqref{k1}, \eqref{B} and \eqref{bAn} we find
\begin{align}
\bP &=  \int_0^\infty  \!\! \int_0^{2\pi} \Ac(k,\theta) k^2 \begin{pmatrix} \cos \theta \\ \sin \theta \end{pmatrix} \, \dd k \dd \theta
= \int_0^{2\pi} \mathcal{P}(\theta)  \begin{pmatrix} \cos \theta \\ \sin \theta \end{pmatrix} \, \dd \theta \nonumber \\
&= \sum_{n=-\infty}^\infty  \int_0^{2\pi} \begin{pmatrix} \cos \theta \\ \sin \theta \end{pmatrix} \, \ee^{n \ii \theta} \, \dd \theta \,p_n = \begin{pmatrix} +\Re  2 \pi  p_1 \\ - \Im  2 \pi  p_1 \end{pmatrix}.
\label{oo}
\end{align}
Substituting \eqref{aa}--\eqref{oo} into \eqref{bL2} yields \eqref{bL3}.

\section{The LHCS spectrum} \label{app:LHCS}

We use the spectrum proposed by \cite{LHCS1963}. This takes the separable form
\ch{
\beq
\bA(k,\theta)= f(k) \times \underbrace{\frac{\Gamma(s+1)}{2 \sqrt{\pi} \Gamma(s+1/2)} \cos^{2s}\left((\theta-\theta_p)/2\right)}_{D(\tha)}. \label{LHCS111}
\eeq
For convenience, we assume that the peak angle $\theta_p = 0$ in this appendix. 
(For $\theta_p \neq 0$, the final expression for $p_n$ should be multiplied by $\ee^{-\ii n \theta_p}$.)
}
In \eqref{LHCS111} the parameter $s \ge 0$ controls the directional spread around the primary direction of wave propagation \ch{conventionally taken along the positive $x$-axis}. The wavenumber function $f(k)$ is chosen so that the frequency spectrum is a truncated Gaussian 
\ch{with standard deviation of  $0.040$ rad s$^{-1}$ and peak angular frequencies at $\sigma = 0.44 $ rad s$^{-1}$ (corresponding to peak period at $14.3$ s) for the Gulf Stream example in figure \ref{fig:hsgulfstream}), or at $\sigma = 0.61$ rad s$^{-1}$ (corresponding to peak period at $10.3$s) for all the other examples. 
}
We take advantage of linearity to report all numerical results in terms of the relative SWH anomaly $\hs/\bHs$ so the amplitude of the Gaussian in \eqref{LHCS111}, proportional to $\bHs$, is unimportant in this paper.

With \eqref{LHCS111}, the function $\Pc(\tha)$ in \eqref{B} becomes
\beq
\Pc (\tha) = \alpha  D(\theta), \quad  \textrm{where} \quad \alpha \defn  \int_0^\infty \!\!\!f(k) k^2 \dd k. 
\label{Pthetas}
\eeq
The Fourier coefficients of $D(\theta)$, denoted   $p_n$ in \eqref{bAn},  are real and satisfy $p_{-n}=p_{n}$. When $s$ is an integer $\Gamma(s+1/2)$ can be expressed  in terms of factorials and then  these coefficients are
\beq
p_n =  \frac{\alpha}{2\pi} \begin{cases}
\displaystyle{\frac{(s!)^2}{(s+n)!(s-n)!}}
 & \textrm{for} \ \ |n| \le s  \\
0 & \textrm{for} \ \ |n| >s.
\end{cases},
\eeq
The wave momentum is
\beq
\bP = \frac{\alpha s}{s+1} \begin{pmatrix} 1 \\ 0\end{pmatrix}.
\eeq
The spectrum \eqref{simpdir} is a particular case of \eqref{LHCS111} with $s=1$.

\section{U2H map for highly directional spectra}\label{app:swell}

We derive an asymptotic approximation for $\hbl(\qang)$  in \eqref{JV7} for a spectrum  \eqref{direc} in the limit $\delta \to 0$. We first note that
\beq
\bP =   \int_{-\pi/\delta}^{\pi/\delta} \mathcal{P}(\Theta) \begin{pmatrix} \cos(\delta \Theta) \\ \sin(\delta \Theta) \end{pmatrix} \, \dd \Theta = 2\pi p_0 \begin{pmatrix} 1 \\ 0 \end{pmatrix} + O(\delta).
\label{asyP}
\eeq 
Thus the first term in \eqref{JV7} makes an $O(1)$ contribution to $\hbl(\qang)$. To evaluate the second term, we approximate the integral involved as
\begin{align}
 \int \frac{\mathcal{P}(\Theta)}{\cos(\delta \Theta-\qang) - \ii \mu} \, \dd \Theta 
 &= \frac{1}{\cos \qang} \int \mathcal{P}(\Theta) \, \dd \Theta + O(\delta) \nonumber \\
 &=\frac{2 \pi p_0}{\cos \qang} + O(\delta).
 \label{asyint1}
\end{align}
This is also $O(1)$ except near $\qang = \pm \pi/2$. There we use the rescaled variables $\Phi_\pm = (\qang \mp \pi/2)/\delta$ to write
\begin{align}
 \int \frac{\mathcal{P}(\Theta)}{\cos(\delta \Theta-\qang) - \ii \mu} \, \dd \Theta 
 &= \pm \int \frac{\mathcal{P}(\Theta)}{\sin(\delta \Theta-\delta \Phi_\pm) \mp \ii \mu} \, \dd \Theta \nonumber \\
 &=\pm \delta^{-1} \int \frac{\mathcal{P}(\Theta)}{\Theta-\Phi_\pm \mp \ii \mu} \, \dd \Theta  + O(\delta).
 \label{asyint2}
\end{align}
Introducing \eqref{asyP}, \eqref{asyint1} and \eqref{asyint2} into \eqref{JV7}, we find 
\beq
\hbl(\qang) \sim \frac{16}{g  \ch{\bHs^2}} \times \begin{cases} 
-\displaystyle{\delta^{-2}  \partial_{\Phi_+} \int_{-\infty}^\infty \frac{\mathcal{P}(\Theta)}{\Theta - \Phi_+ - \ii \mu} \, \dd \Theta \,  \, \eqp+O(1)}  &\textrm{for} \ \ |\qang - \pi/2| = O(\delta) \\
\displaystyle{\delta^{-2}  \partial_{\Phi_-} \int_{-\infty}^\infty \frac{\mathcal{P}(\Theta)}{\Theta - \Phi_- + \ii \mu} \, \dd \Theta \,  \, \eqp+O(1)}  &\textrm{for} \ \ |\qang + \pi/2| = O(\delta) \\
  \displaystyle{-2 \pi p_0 \left(\frac{\sin \qang}{\cos^2 \qang} \, \eqp + \begin{pmatrix} 2 \\ 0 \end{pmatrix} \right)}& \textrm{for} \ \ |\qang \mp \pi/2| = O(1)
\end{cases}. 
\label{asybL}
\eeq

Under the assumption of a localised current, 
we now derive an approximation for $\hs(\bx)$ by introducing \eqref{asybL} in the inverse Fourier transform 
\beq
\hs(\bx) \ch{/\bHs}= \frac{1}{(2\pi)^2}\int \hbl(\qang) \bcdot \hbU(\bq) \ee^{\ii \bq \bcdot \bx} \, \dd \bq
\label{hsxft}
\eeq
of \eqref{linmap}. We compute explicitly the contribution of the boundary layer at $\qang=\pi/2$; the contribution of the other boundary layer is its complex conjugate, denoted by $\mathrm{c.c.}$, and is added. Using the polar representation of $\bq$ and the approximation
\beq
\bq = q (-\delta \Phi, 1) + O(\delta^2),
\eeq
where we denote $\Phi_+$ by $\Phi$ for simplicity, we obtain
\begin{align}
\hs(\bx) \sim& \frac{16\delta^{-1}}{(2\pi)^2 g \bHs} \int_0^\infty \hat U(0,q) q \, \ee^{\ii q y} \, \dd q \int \left( \dt{}{\Phi} \int \frac{\mathcal{P}(\Theta)}{\Theta-\Phi - \ii \mu} \, \dd \Theta \right) \, \ee^{-\ii \delta q x \Phi} \, \dd \Phi \nonumber \\
&+ \mathrm{c.c.},
\label{long}
\end{align}
Here we implicitly assume that $x = O(\delta^{-1})$ since this turns out to be the range of $x$ for which $\hs$ is the largest.
Integrating by parts in $\Phi$ and swapping  the order of the $\Theta$ and $\Phi$  integrations produces
\beq
\hs(\bx) \sim \frac{16 \ii  x}{(2 \pi)^2 g \bHs} 
\int_0^\infty \hat U(0,q) q^2  \ee^{\ii q y} \, \dd q \int \mathcal{P}(\Theta) \left( \int \frac{ \dd \Phi}{\Theta-\Phi - \ii \mu}    \, \ee^{-\ii q \delta x \Phi} \right) \, \dd \Theta  + \mathrm{c.c.} 
\label{verylong}
\eeq
The $\Phi$-integral   can be evaluated as a contour integral. For $x \ge 0$, we close the contour in the lower half plane where the simple pole $\Phi = \Theta - \ii \mu$ is located. In the limit $\mu \to 0^+$ the integral becomes $2 \pi \ii \ee^{-\ii q \delta x \Theta}$, leading to
\beq
\hs(\bx) \sim - \frac{8x}{\pi g \bHs} 
\int_0^\infty \hat U(0,q) q^2 \ee^{\ii q y} \, \dd q \int \mathcal{P}(\Theta) \ee^{-\ii q \delta x\Theta} \, \dd \Theta  + \mathrm{c.c.}.
\eeq
Recognising the integral with respect to $\Theta$ as the Fourier transform $\hat{\mathcal{P}}(q\delta x)$ of $\mathcal{P}(\Theta)$, we obtain the compact expression
\beq
\hs(\bx) \sim - \frac{8x}{\pi g \bHs} 
\int_0^\infty \hat U(0,q) \hat{\mathcal{P}}(q \delta x) q^2 \ee^{\ii q y} \, \dd q   + \mathrm{c.c.}  \quad \textrm{for} \ \ x \ge 0.
\eeq
For $x<0$, the integration contour can be closed in the upper half $\Phi$-plane, where there are no poles, hence $\hs(\bx)=0$ for $x<0$. The reality conditions $\hat U^*(0,q)=\hat U(0,-q)$ and $\hat{\mathcal{P}}^*(q \delta x) = \hat{\mathcal{P}}(-q \delta x)$ can then be used to reduce the result to:
\beq
\hs(\bx) \sim - \frac{8 x}{\pi g \bHs} 
\int_{-\infty}^\infty \hat U(0,q) \hat{\mathcal{P}}(q \delta x) q^2 \ee^{\ii q y} \, \dd q \quad \textrm{for} \ \ x \ge 0,
\label{hsasy}
\eeq
and $\hs(\bx)=0$ for $x<0$.

We now show that \eqref{hsasy} is equivalent to the result obtained by \citet{WVBYV} in the linear regime $\eps \ll \delta \ll 1$. In the notation of the present paper, their result reads
\beq
h_s(\bx)=-\frac{\bHs}{2} \int D'(\Theta)\Delta(y-x\Theta)\dd\Theta \quad \textrm{for} \ \ x \ge 0,
\label{hss}
\eeq
where $D(\Theta)$ is the (normalised) angular distribution of the spectrum and
\beq \label{defldef}
 \Delta(y) \defn \frac{1}{\ch{c_{g\star}}  }\int_{-\infty}^ \infty
 \!\!\! \zeta(x,y)\, \dd x, 
 \eeq
with $\ch{c_{g\star}}$ the peak group speed. The function $\Delta(y)$ can be written in terms of the Fourier transform of the current as
\beq
\Delta(y)= -\frac{\ii}{2\pi \ch{c_{g\star}}}\int q_2 \hat U(0,q_2)\ee^{\ii q_2 y} \, \dd q_2.
\label{Del3}
\eeq
Introducing \eqref{Del3}  into \eqref{hss} 
\beq
 \hs =-\frac{\bHs x}{4\pi \ch{c_{g\star}}} \int \hat U(0,q)  \hat D(qx) q^2  \ee^{\ii q y}\dd q\quad \textrm{for} \ \ x \ge 0.  \label{hsswell2}
\eeq
The result in \eqref{hsswell2} is shown to be identical to \eqref{hsasy} by noting that
 \begin{align}
 \bHs^2 D(\theta) &= 16g^{-1} \iint \sigma(k)\mathcal{\bA}k \dd k \dd \theta \, D(\theta) 
 = 32g^{-1} \iint \ch{c_{g}}\mathcal{\bA}k^2 \dd k \dd \theta  \\
 & \approx 32 g^{-1}\ch{c_{g\star}} \mathcal{P}(\theta).
 \end{align}
Here we use the definition \eqref{B} of $\mathcal{P}(\theta)$ in the case of a separable spectrum and we approximate $\ch{c_{g}}(k) \approx \ch{c_{g\star}}$ as appropriate for a highly directional spectrum. 

\section{Axisymmetric vortex} \label{app:GVhs} 
For a purely vortical, axisymmetric flow, with $\hat \phi(q) = 0$ and $\hat \psi(q) = - q^{-2} \hat \zeta(q)$, 
combining \eqref{Lp17} into  \eqref{helmhs} yields
\beq
\hs(\bq) =-\frac{16 \ii}{g \bHs}  \sum_{n=-\infty}^\infty n (-\ii)^{|n|} \, 2 \pi \tilde p_n \ee^{ \ii n \qang} \,  q^{-1}  \hat \zeta (q)\per
\eeq
The  inverse Fourier transform is
\beq
\hs(\bx)  =  -\frac{16 \ii}{g \bHs}  \sum_{n=-\infty}^\infty n (-\ii)^{|n|} \, \frac{2 \pi \tilde p_n}{2 \pi}  \,\int_0^\infty  \!\!\!\hat   \zeta(q) \int_0^{2\pi} \ee^{\ii  qr \cos (\qang - \xang) + \ii n \qang}\frac{ \dd \qang}{2 \pi}\, \dd q \,\per
\label{IFT3}
\eeq
In the $\qang$-integral above,  $\xang$ is the  polar angle in physical space. The $\qang$-integral can be reduced to Bessel functions
\begin{align}
 \int_0^{2\pi} \ee^{\ii  qr \cos (\qang - \xang) + \ii n \qang} \frac{ \dd \qang}{2 \pi} 
 &=  \int_0^{2\pi} \ee^{\ii  qr \cos \alpha } \cos( n \alpha) \, \frac{ \dd \alpha}{2 \pi}\, \ee^{\ii n \xang}  \\
 &= \ii^{|n|} J_{|n|}(qr) \ee^{\ii n \xang}\per \label{Bess1}
\end{align}
Substituting \eqref{Bess1} into  \eqref{IFT3} we obtain
\beq
\hs(\bx)  =  - \frac{16 \ii}{g \bHs}  \sum_{n=-\infty}^\infty n  \tilde p_n \,\int_0^\infty  \!\!\!\hat   \zeta(q)J_{|n|}(qr)  \, \dd q \,\ee^{\ii n \xang},
\label{IFT7}
\eeq
valid for any vortical axisymmetric flow.
%
For the Gaussian vortex in \eqref{Gzeta} the $q$-integral reduces to
\begin{align}
\int_0^{\infty}\!\!\! J_{|n|}(qr)  \hat \zeta(q) \,\dd q &=   \kappa \int_0^{\infty}\!\!\! J_{|n|}(qr)   \ee^{-r_{\mathit{v}}^2 q^2/2}\,\dd q \\
&=\frac{\kappa}{r_v}\sqrt{\frac{\pi}{2}}\ee^{-r^2/4 r_v^2}I_{|n|/2}(r^2/4 r_v^2),
\end{align}
where $I_{|n|/2}$ is the  modified Bessel function. This simplifies  $\hs(\bx)$ in \eqref{IFT7}  to  \eqref{Bess7}.

\bibliographystyle{jfm}
\bibliography{currentSWH_finalJFM}
\end{document}